\documentclass[a4paper,12pt]{amsart}
\usepackage{amsfonts,amsmath,amssymb,amsthm}
\usepackage[margin=3cm]{geometry}
\usepackage{graphicx}
\usepackage{subfigure}
\usepackage{float}
  \restylefloat{figure}
\usepackage{mathrsfs}
\usepackage{tikz}
  \usetikzlibrary{shapes}
  \usetikzlibrary{snakes}

\title{On many body system interactions}
\author{Nils A.\ Baas}
\address{Department of Mathematical Sciences, NTNU, NO-7491 Trondheim,
  Norway}
\email{baas@math.ntnu.no}
\date{November 1, 2010}

\newcommand{\R}{\mathbb{R}}
\newcommand{\calp}{\mathcal{P}}

\theoremstyle{plain}
\newtheorem*{theorem*}{Theorem}

\theoremstyle{definition}
\newtheorem*{definition*}{Definition}
\newtheorem*{example*}{Example}

\begin{document}
\begin{abstract}
  We discuss possible relationships between geometric and topological
  interactions on one side and physical interactions on the other
  side.  This paper is a follow up of \cite{NewStates}.
\end{abstract}

\maketitle

We will here discuss possible interactions of families of particles.
By a particle we mean a system or an (extended) object in some space.
The families may be finite, countable or uncountably infinite.

\begin{equation*}
  \calp = \{P_i\}_{i \in I}
\end{equation*}

What does it mean that the particles interact?

Often we have ``state''-spaces associated to the particles
\begin{equation*}
  P_i \mapsto S_i.
\end{equation*}
An interaction is a rule telling us how the states influence each other
or are related:
\begin{equation*}
  R \subset \prod S_i
\end{equation*}
If this is stable and time independent we may call it a \emph{bond}.\\

\noindent \textbf{Simplicial model:}\\

A \emph{pair} interaction is geometrically represented as:
\raisebox{-0.55cm}[0pt][0pt]{\begin{tikzpicture}
  \draw[thick,font=\scriptsize]
    (0,0) node[below,yshift=-0.10cm]{$P_i$} -- (2,0)
    node[below,yshift=-0.10cm]{$P_j$};
  \draw[thick,fill=black] (0,0) circle(0.10cm);
  \draw[thick,fill=black] (2,0) circle(0.10cm);
\end{tikzpicture}}\\

An $n$-tuple interaction may be represented as an $n$-simplex, but it
is reducible to pair interactions:

\begin{center}
  \begin{tikzpicture}[thick]
    \draw (0,0) -- (3,-1) -- (5,1) -- (2.5,3) -- (0,0);
    \draw (2.5,3) -- (3,-1);
    \draw[dashed] (0,0) -- (5,1);
    \draw (0,0) node[left]{$P_1$};
    \draw (3,-1) node[below]{$P_2$};
    \draw (5,1) node[right]{$P_3$};
    \draw (2.5,3) node[above]{$P_4$};
  \end{tikzpicture}
\end{center}

Here we represent the systems geometrically by \emph{points} (in some
Euclidean space):

\begin{equation*}
  \text{Particle (system)} \quad \longmapsto \quad \text{point in f.\
    ex.\ $\R^3$}
\end{equation*}

But we may have more sophisticated interactions like:\\

%\newpage

\noindent \textbf{Brunnian or Borromean model:}\\

$n$ particles interact in such a way that no subset of them interact.
This suggest and is best understood by another representation:

\begin{equation*}
  \text{Particle} \quad \longmapsto \quad \text{ring (string) in
    $\R^3$}
\end{equation*}

\begin{figure}[H]
  \centering
  \subfigure[Borromean]{
    \includegraphics{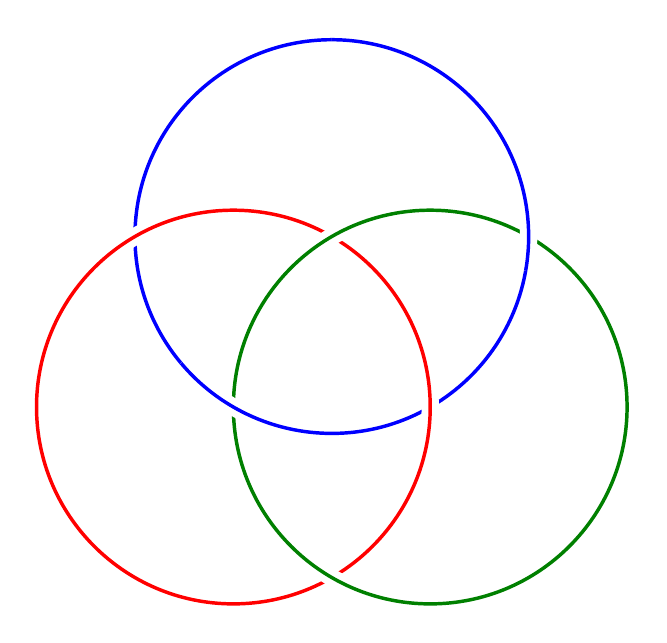}
  } \qquad \qquad
  \subfigure[Brunnain interactions of length $k = 4$]{
    \includegraphics[scale=0.25]{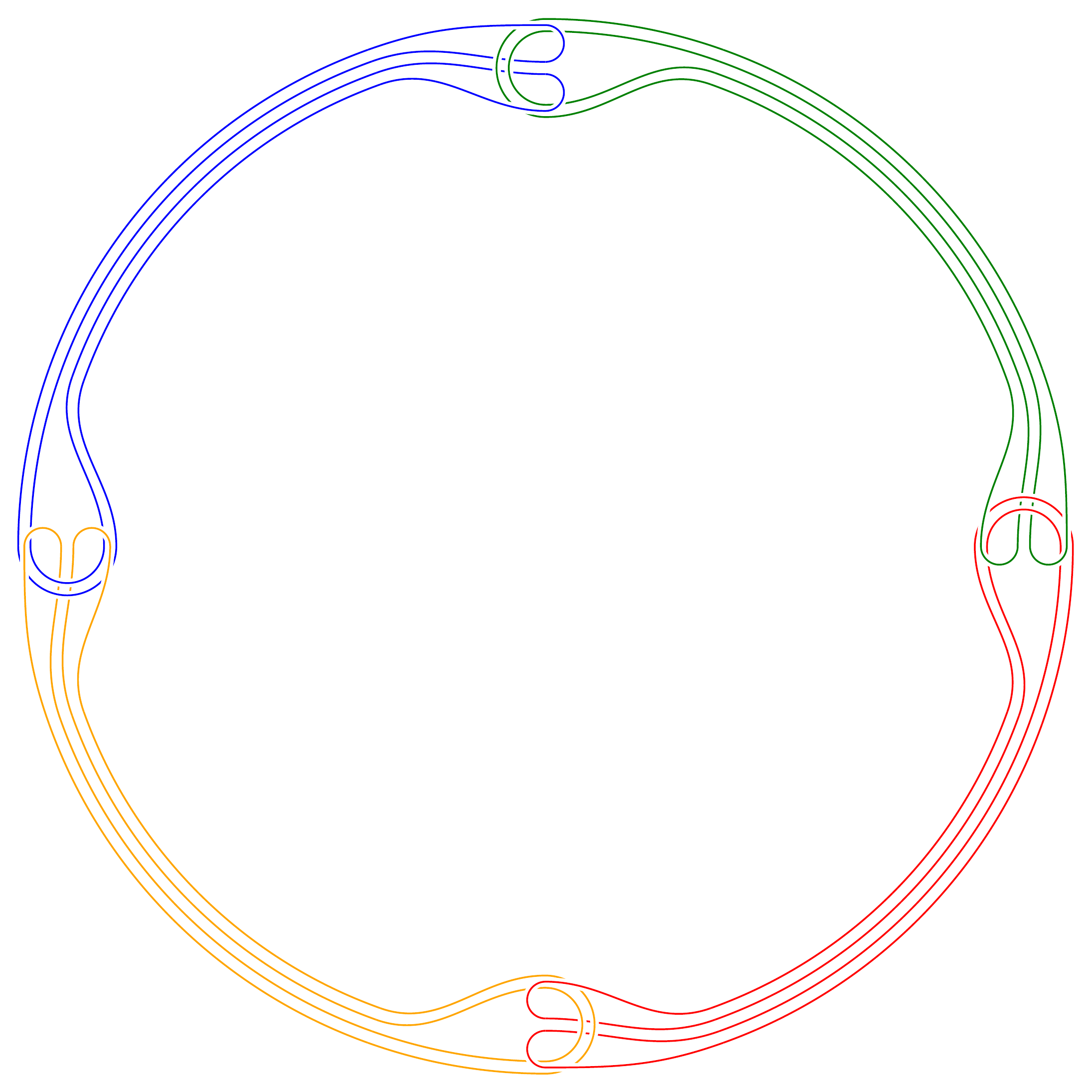}
  }
\end{figure}

In this representation a pair interaction is represented by Hopf
links:

\begin{figure}[H]
  \centering
  \subfigure{
    \includegraphics[scale=0.75]{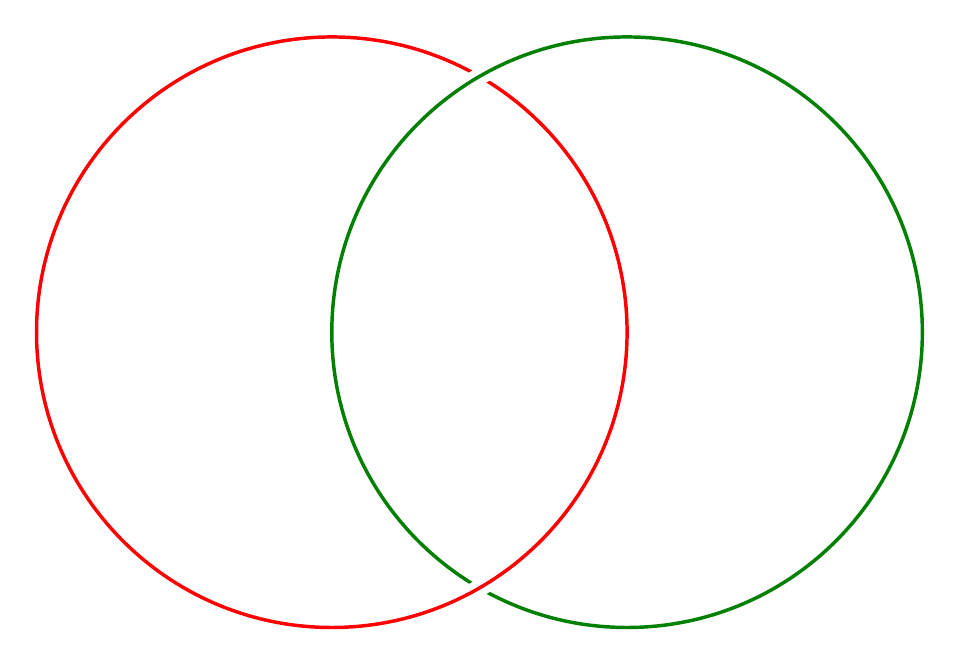}
  } \qquad \raisebox{2.5cm}[0pt][0pt]{or} \qquad
  \subfigure{
    \includegraphics[scale=0.225]{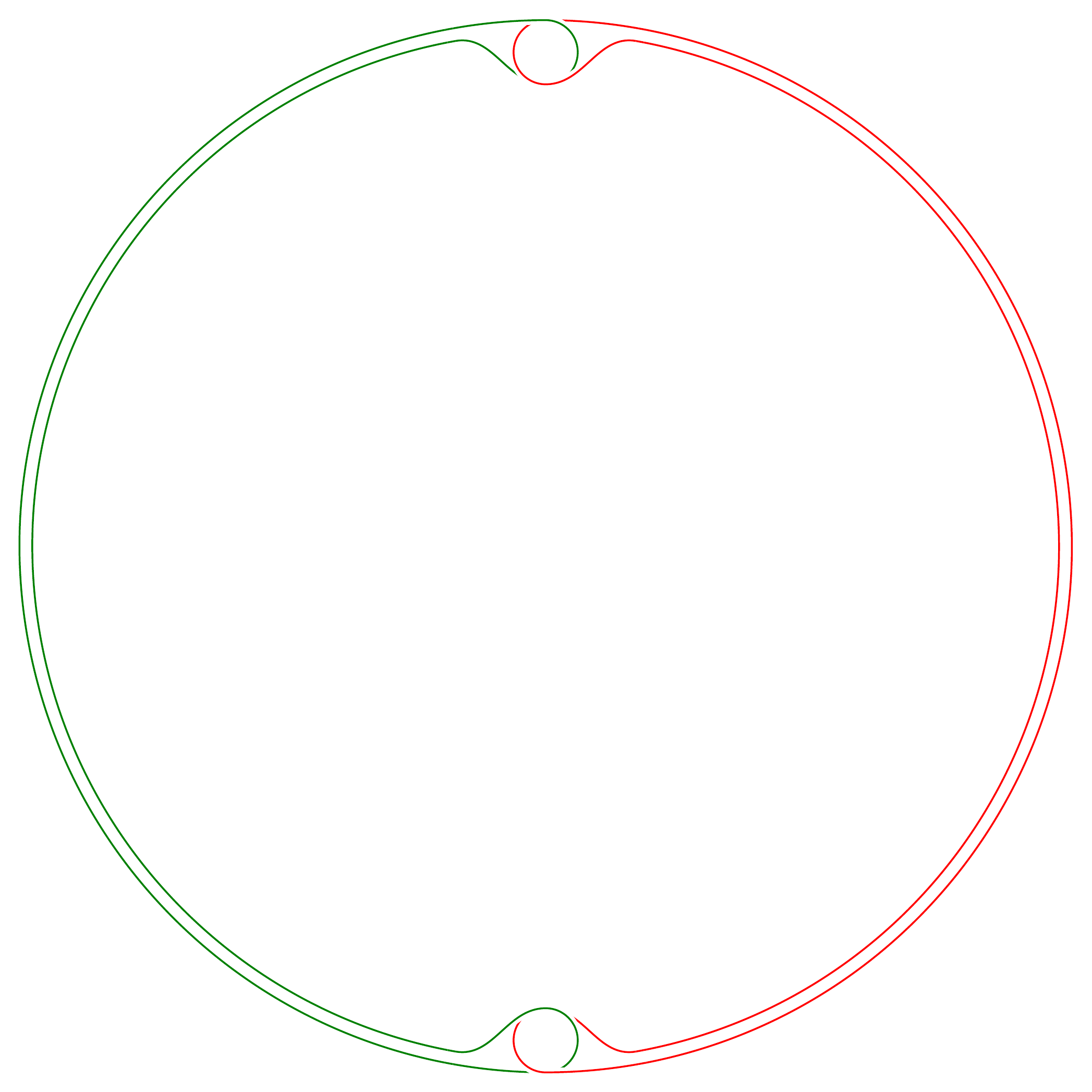}
  }
\end{figure}

With this representation we have in \cite{NewStates} introduced a
whole new hierarchy of (possible) higher order interactions
represented by new higher order links of rings.  This has been
extensively developed in \cite{NewStates}.\\

The interesting question is then:\\

\emph{What about other geometric (topological) representations, and
  what kind of new interactions do they suggest --- both of first and
  higher order?}\\

We may proceed as follows.

\begin{equation*}
  \text{Particle} \quad \longmapsto \quad \text{Space (of some kind
    and f.\ ex.\ embedded in a fixed ambient space $A$)}
\end{equation*}

\begin{equation*}
  P_i \longrightarrow T_i = \mathrm{Space}_i \subseteq A
\end{equation*}

Pictorially this looks like

\begin{figure}[H]
  \centering
  \includegraphics{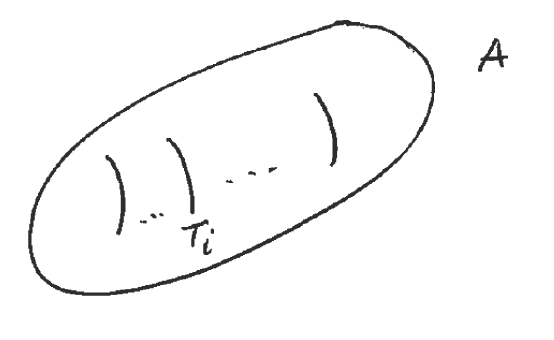}
\end{figure}

In the sense of \cite{Hyper} $A$ is a bond of $\{T_i\}$.  This can be
iterated and one may form higher order bonds, ending up with a
\emph{hyperstructure}, defined and described in \cite{Hyper}.

Let us be more specific and consider the situation where the
representing spaces are manifolds embedded in a large ambient manifold
or Euclidean space.

\begin{equation*}
  P_i \longrightarrow M_i \subseteq B (\subseteq \R^N)
\end{equation*}

\vspace*{2cm}

\begin{figure}[H]
  \centering
  \includegraphics[scale=1.25]{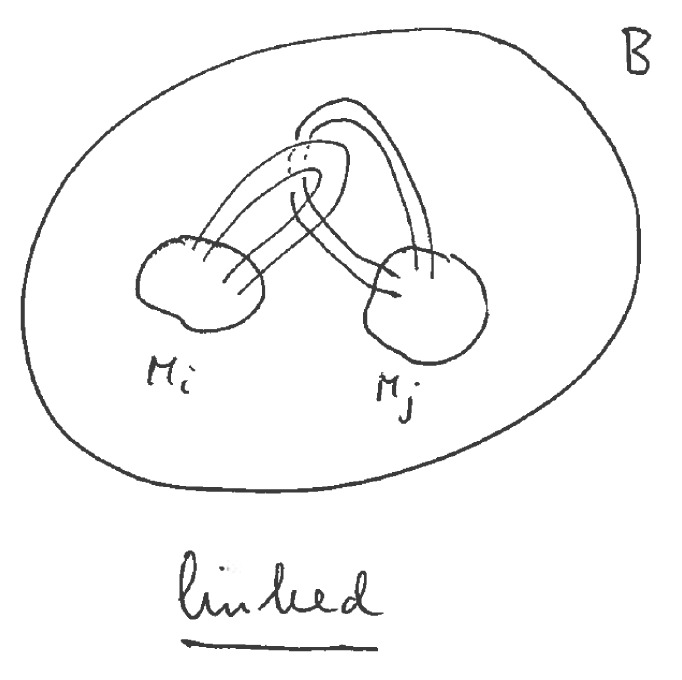}
\end{figure}

\newpage

\vspace*{3cm}

\begin{figure}[H]
  \centering
  \includegraphics[scale=1.5]{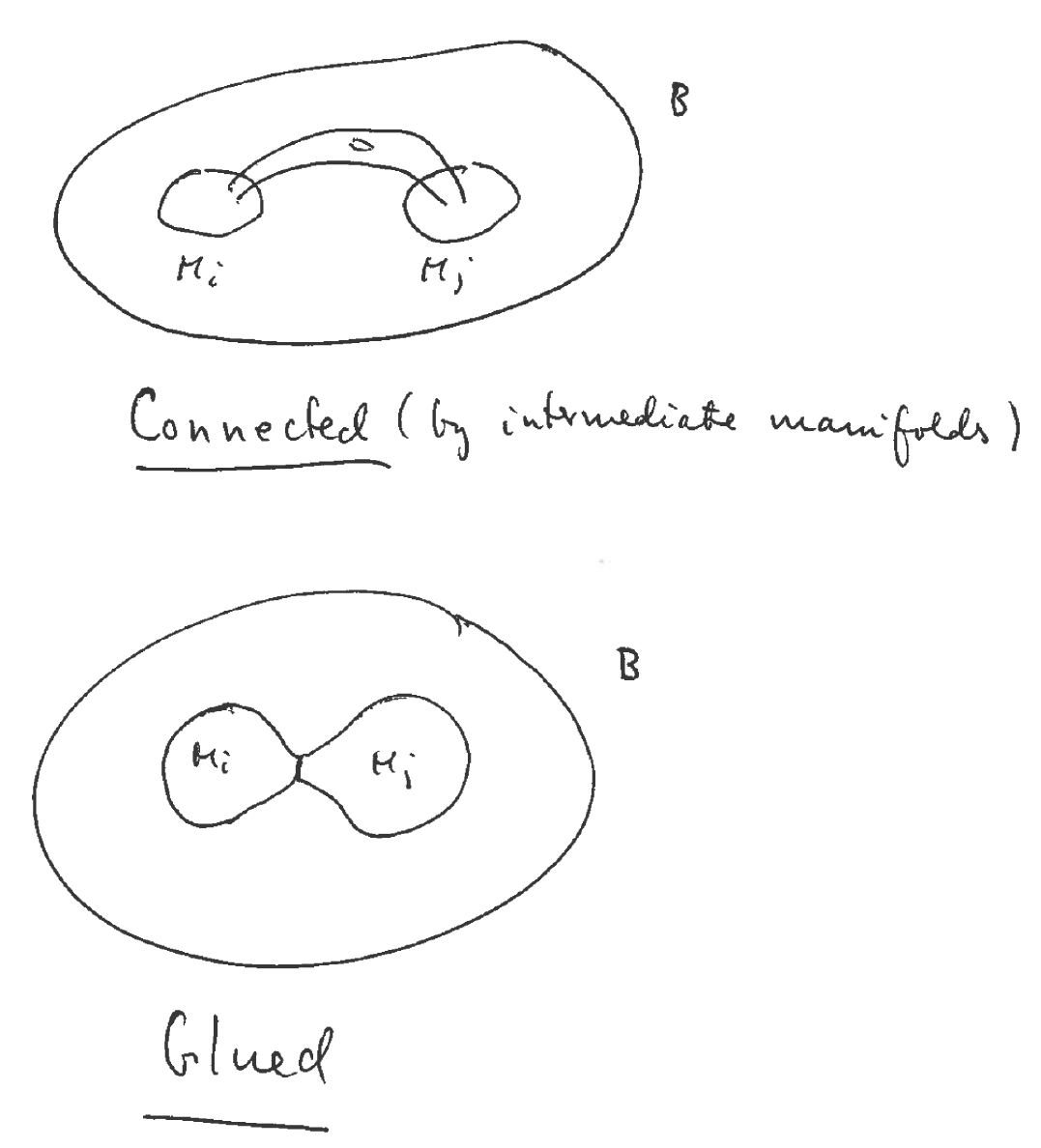}
\end{figure}

\newpage

\begin{figure}[H]
  \centering
  \includegraphics[scale=1.5]{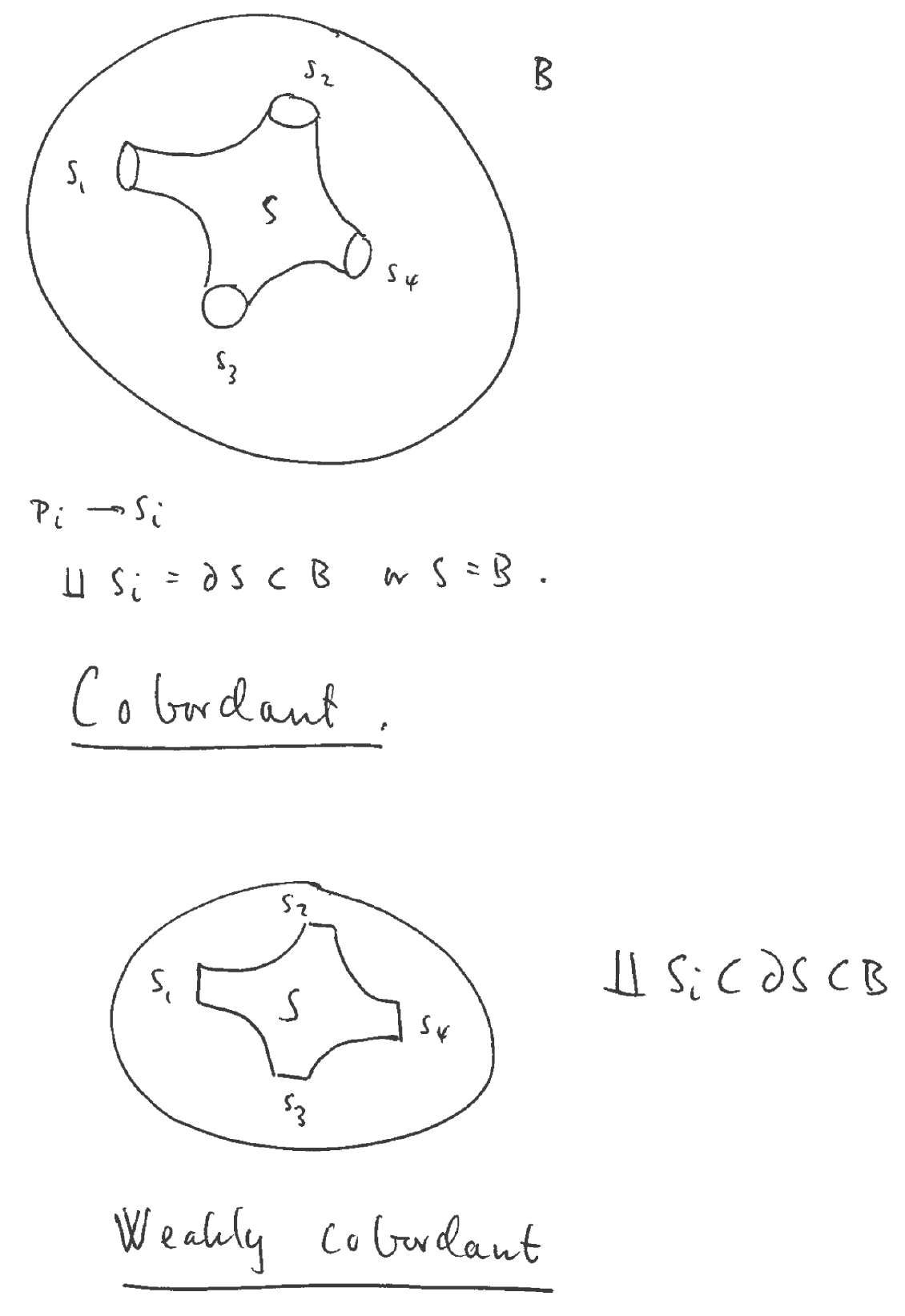}
\end{figure}

\newpage

\begin{example*} $B = \R^3$
  \begin{figure}[H]
    \centering
    \begin{tikzpicture}
      \draw[thick] (0,0) ellipse (3cm and 2cm);
      \draw (2,0) node{$T_i$};
      \draw (3,1.8) node{$B$};
    \end{tikzpicture}
    \hspace*{-5.5cm} \raisebox{0.5cm}[0pt][0pt]{
      \includegraphics[scale=0.5]{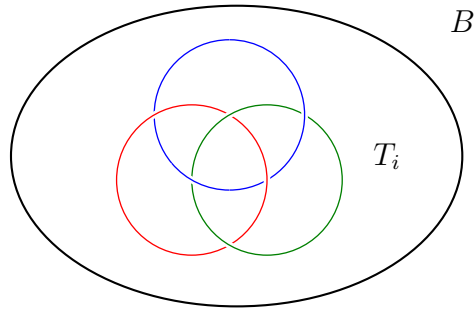}}
    \caption{Borromean interactions}
    \label{fig:BorroInt}
  \end{figure}

  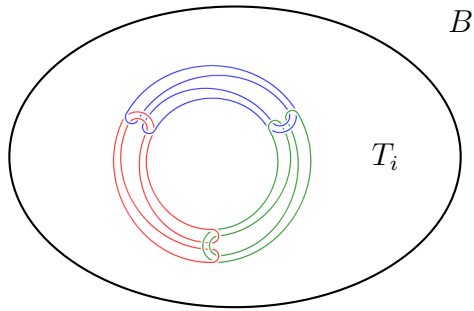
\begin{figure}[H]
    \centering
    \begin{tikzpicture}
      \draw[thick] (0,0) ellipse (3cm and 2cm);
      \draw (2,0) node{$T_i$};
      \draw (3,1.8) node{$B$};
    \end{tikzpicture}
    \hspace*{-5.5cm}
    \raisebox{0.5cm}[0pt][0pt]{
      \includegraphics[scale=0.35]{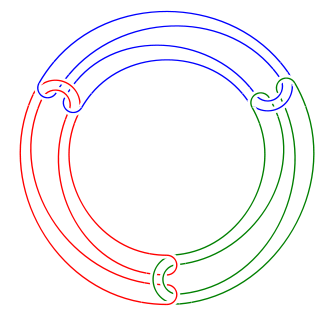}}
    \caption{Brunnain interactions}
    \label{fig:BrunnInt}
  \end{figure}

  See \cite{NewStates} for a whole hierarchy of extensions.
\end{example*}

For higher dimensional manifolds there is a variety of ways to do
this (spaces of embeddings).

One may manipulate or tune externally a system in such a way that it
is represented by a desired (linked,$\ldots$) embedding of the $M_i$'s
in $B$.

Hence it justifies calling $B$ a bond, see \cite{Hyper}.\\

Then one may --- as already mentioned --- iterate this process to
higher order bonds which we have defined as hyperstructures.  This
gives higher order, hyperstructured interaction patterns of the
particles.\\

\emph{This means that for many body systems there is a whole new
universe of new represented states}.

\emph{The pertinent question is then: Which of these new types of
  states are realizable in real world systems (physical, chemical,
  biological, social,$\ldots$)?}\\

This discussion also shows that bonds of subspaces (like manifolds)
and their associated hyperstructures may be a good laboratory for
suggesting new states, designing and studying general many body
systems and their interactions.  But the geometric interactions in the
geometric universe should then be interpreted back into interactions
in the real universe where the particles live.  In cold gases
Borromean or Brunnian states of first order correspond to Efimov
states, see \cite{NewStates} for details.

In \cite{NewStates} we have studied and suggested connections between
physical states and higher order links in the geometric universe of
links.

\nocite*


\begin{thebibliography}{1}
\bibitem{Hyper} N.A.\ Baas, Hyperstructures as Abstract Matter, Adv.\
  in Complex Systems
\bibitem{NewStates} N.A.\ Baas, New States of Matter, Preprint, NTNU,
  2010
\end{thebibliography}
\end{document}